\documentclass[aps,prl,nofootinbib,nobibnotes,twocolumn,floatfix,letterpaper,superscriptaddress,showpacs]{revtex4}
\usepackage{graphicx}\usepackage{natbib}\usepackage{amsmath}
\pdfoutput=1
\usepackage{epstopdf}
\begin{document}

\title{The Cosmic MeV Neutrino Background as a Laboratory for Black Hole Formation}

\author{Hasan Y{\"u}ksel}
\affiliation{Theoretical Division, Los Alamos National Laboratory, Los Alamos, NM 87544}
\affiliation{Lawrence Berkeley National Laboratory, Berkeley, CA 94720}
\affiliation{Department of Physics, Mimar Sinan Fine Arts University, Bomonti 34380, \.{I}stanbul, Turkey}

\author{Matthew D. Kistler}
\affiliation{Lawrence Berkeley National Laboratory, Berkeley, CA 94720}
\affiliation{Department of Physics, University of California, Berkeley, CA 94720}
\affiliation{Einstein Fellow}

\date{October 2, 2014}

\begin{abstract}
Calculations of the cosmic rate of core collapses, and the associated neutrino flux, commonly assume that a fixed fraction of massive stars collapse to black holes.  We argue that recent results suggest that this fraction instead increases with redshift.  With relatively more stars vanishing as ``unnovae'' in the distant universe, the detectability of the cosmic MeV neutrino background is improved due to their hotter neutrino spectrum, and expectations for supernova surveys are reduced.  We conclude that neutrino detectors, after the flux from normal SNe is isolated via either improved modeling or the next Galactic SN, can probe the conditions and history of black hole formation.
\end{abstract}


\pacs{97.60.Bw, 98.70.Vc, 95.85.Ry, 14.60.Pq}
\maketitle

\begin{figure}[b!]
\includegraphics[width=3.35in,clip=true]{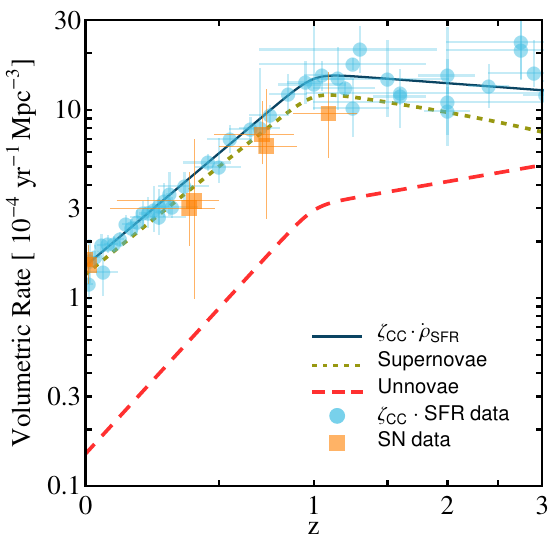}
\vspace{-0.2cm}
\caption{The cosmic rate of core collapse.  Shown are recent measurements of core-collapse supernovae \cite{Botticella:2011nd,Melinder:2012nv,Mattila:2012zr,Dahlen:2012cm} ({\it squares}; see \cite{Horiuchi:2011zz} for older data), which fall just below the expectation from star formation rate data with all stars of mass $\gtrsim\,$8$\,M_\odot$ yielding optical SNe ({\it circles}; \cite{Hopkins:2006bw}).  These are compared to our model assuming a local 10\% rate of unnovae that evolves with $z$ ({\it dashed}), the predicted SN rate ({\it dotted}), and the total ({\it solid}).
}
\label{rates}
\end{figure}

{\bf Introduction.---}
The dearth of supernovae in our own galaxy leads us to examine those that occur throughout the universe in order to study the physics underlying the collapse of short-lived massive stars, which is vital for understanding stellar life and death \cite{Baade,Colgate:1966ax,Raffelt:1996wa,Arnett:1996ev,Heger:2002by,Marek:2008qi,Yakunin:2010fn,Ott:2012kr}.  Observations of the classes of SNe attributed to core collapse -- Types II, Ib, and Ic -- have advanced greatly in the past decade, and the most recent measurements of their rates now cover out to a redshift of $z$$\, \gtrsim\,$1 \cite{Botticella:2011nd,Melinder:2012nv,Mattila:2012zr,Dahlen:2012cm,Li:2010kd,Graur:2011cv,Bazin:2009mp,Cappellaro:1999qy}.  In Fig.~\ref{rates}, we see that these data are close to, yet do not quite match \cite{Horiuchi:2011zz}, the assumption that all $\gtrsim\,$8$\,M_\odot$ stars explode as SNe \cite{Hopkins:2006bw}.

However, some subset of core collapses must result in the stellar-mass black holes seen in the Milky Way and beyond \cite{Fender:2012tx,Mirabel:2003st}.  Despite many years of research \cite{Oppenheimer:1939ue,Wilson(1971),Fryer:1999mi,Zhang:2007nw,O'Connor:2010tk,Ugliano:2012kq,Horiuchi:2014ska}, the fraction that do so remains uncomfortably uncertain.  One option is to search for stars in nearby galaxies that simply disappear, i.e., unnovae (UNe) \cite{Kochanek:2008mp}. Throughout, we generically refer to collapses yielding a neutron star and bright optical transient as ``SNe'' \cite{Zwicky:1940zz},
``unnovae'' as those leading to a black hole (which may also
include some type of photon emission).

Fortunately, even if no photons result from their core collapse, stars do not vanish entirely without a trace.  Such massive progenitors yield, if only for an abbreviated period, protoneutron stars that emit copious amounts of neutrinos, with a hotter spectrum than in a lower-mass collapse \cite{Burrows:1988ba,Baumgarte:1996iu,Sumiyoshi:2007pp,Sumiyoshi:2008zw,Nakazato:2008vj,Fischer:2008rh,Nakazato:2012qf,Nakazato:2015rya}.  Thus, in addition to the diffuse supernova neutrino background (DSNB) from successful explosions (\cite{ZG,Hartmann:1997qe,Kaplinghat:1999xi,Ando:2002ky,Strigari:2005hu,Yuksel:2005ae,Yuksel:2007mn,Horiuchi:2008jz}; see \cite{Beacom:2010kk,Lunardini:2010ab} for a comprehensive review), unnovae should contribute to the overall cosmic MeV neutrino background (CMNB) (e.g., \cite{Lunardini:2009ya,Lien:2010yb,Keehn:2010pn}).

In contrast to prior CMNB studies, which assumed that a uniform fraction of core collapses result in unnovae throughout cosmic history, we argue that this fraction is instead larger in the more distant universe than locally.  This is because lower stellar metallicity points toward a greater propensity for black hole formation \cite{Zhang:2007nw,O'Connor:2010tk} and the metallicity of star forming gas was lower at higher redshift \cite{Tremonti:2004et,Kewley,Erb:2006qy,Savaglio:2005hi,Maiolino:2008gh}.  Although a first-principles model of the cosmic unnova rate is not yet available, we draw guidance from gamma-ray bursts (GRBs), which in the collapsar model arise from core collapses that yield rapidly-rotating black holes (\cite{Woosley:1993wj}; cf.\ \cite{Metzger:2010pp}).  GRB observations indeed show a sensitivity to metallicity (e.g., \cite{Stanek:2006gc,Graham:2012ga}), in accord with theory \cite{Yoon:2005tv,Woosley:2005gy}, and a stronger evolution with redshift than the star formation rate (SFR) \cite{Daigne:2006kf,Le:2006pt,Yuksel:2006qb,Kistler:2007ud,Kistler:2009mv,Butler:2009nx,Kistler:2013jza}.

Using cosmic GRB data as an empirical proxy for the changing rate of black hole formation, in combination with core-collapse neutrino simulations from \cite{Nakazato:2012qf} and modern SFR measurements, we find that accounting for this evolution is crucial, with unnovae plausibly forming the dominant CMNB component.  This approach leads to qualitatively-different implications, including that the CMNB can provide a powerful near-term probe of the physics of black hole formation {\it even if astronomical observations would suggest otherwise}.  In particular, models assuming a constant UN fraction would naturally be normalized to data from surveys looking for disappearing stars \cite{Kochanek:2008mp} or associated faint transients \cite{Lovegrove:2013ssa,Piro:2013voa}.  However, since these are limited to the metal-enriched local universe where the UN rate is lowest (possibly even below their sensitivities), this could be dangerously misleading.

As experiments near the expected level of the CMNB \cite{Malek:2002ns,Eguchi:2003gg,Aharmim:2006wq,Bays:2011si}, an approach as presented here is needed to avoid misinterpreting the eventual discovery.  We address the capabilities of next-generation detectors (e.g., \cite{Abe:2011ts,Goon:2012if,Autiero:2007zj}) to extract properties of neutrino emission from black hole formation.  Astronomical data, such as from LSST \cite{Abell:2009aa}, can test our expectation of the SN rate not simply scaling from the SFR (which may already be hinted at in Fig.~\ref{rates}).

\begin{figure}[t!]
\includegraphics[width=3.35in,clip=true]{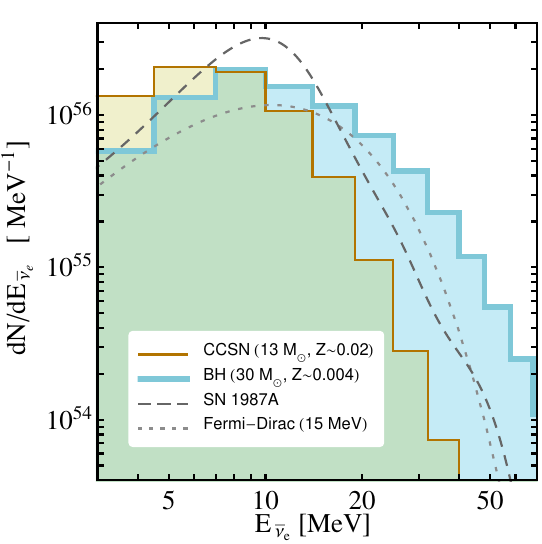}
\caption{The $\bar{\nu}_e$ spectra used in this study.  Shown are those from the 13$\,M_\odot$ ($Z$$\,=\,$0.02, 100~msec revival) SN simulation of Nakazato et al.\ ({\it thin solid}), their 30~$M_\odot$ model yielding a black hole ({\it thick solid}) \cite{Nakazato:2012qf}, and the SN~1987A model of \cite{Yuksel:2007mn} ({\it dashed}).  These are compared to a Fermi-Dirac spectrum with $\langle E_{\bar{\nu}_e}  \rangle=\,$15~MeV and  $\mathcal{L}_{\bar{\nu}_e}= 5 \times 10^{52}$ erg ({\it dotted}).
}
\label{fluxes}
\end{figure}

{\bf Neutrino spectra from core collapse.---}
For water Cherenkov detectors, the principal detection channel of the CMNB is inverse beta decay, $\bar\nu_e+p\to n+e^+$, so our primary interest is in the total $\bar\nu_e$ flux arriving at Earth from distant core collapses.  We consider two scenarios for the SN contribution.  The first takes the time-integrated spectrum from SN~1987A data \cite{Hirata:1987hu,Bionta:1987qt}, as inferred in \cite{Yuksel:2007mn}, as representative of all SNe.  This spectrum, shown in Fig.~\ref{fluxes}, has $\langle E_{\bar\nu_e}  \rangle$$\,=\,$12~MeV and $\mathcal{L}_{\bar\nu_e}$$\,=\,$$ 6 \times 10^{52}\,$erg.  This has the advantage of naturally including any oscillation effects on the outgoing spectrum, but suffers from sampling only one star with limited statistics.  For comparison, we also display a Fermi-Dirac spectrum with $\langle E_{\bar\nu_e}  \rangle$$\,=\,$15~MeV and  $\mathcal{L}_{\bar\nu_e}$$\,=\,$$ 5 \times 10^{52}\,$erg, as is often used.

As an alternative, we consider the results of Nakazato et al.\ \cite{Nakazato:2012qf}, who combined general relativistic radiation hydrodynamical simulations, assuming shock revival at either 100, 200, or 300~msec after bounce, and protoneutron star cooling until 20~sec to find neutrino light curves and spectra for four progenitor masses (13, 20, 30, and 50$\,M_\odot$) at two metallicities ($Z$$\,=\,$0.02 or 0.004).  We show the time-integrated $\bar\nu_e$ spectrum for 13$\,M_\odot$ and $Z$$\,=\,$0.02 in Fig.~\ref{fluxes}, using the 100~msec model to be conservative.  Convolving the models over a Salpeter mass function yields a very similar spectrum. Since other flavors have similar spectra, modifications due to neutrino mixing or neutrino-neutrino interactions  (e.g.~\cite{Chakraborty:2008zp,Galais:2009wi,Duan:2010bg,Lunardini:2012ne}) should be small, so we use this spectrum in determining the DSNB.

Nakazato et al.\ found that their 30$\,M_\odot$, $Z$$\,=\,$0.004 model yielded a black hole.  The time-integrated $\bar\nu_e$ spectrum \cite{Nakazato:2012qf} is shown in Fig.~\ref{fluxes} and is far harder than from SNe.  We will use this as the template for the unnova contribution to the CMNB.  In general, the flux from black hole production will depend on the progenitor, the nuclear equation of state and explosion mechanism \cite{Sumiyoshi:2007pp,Sumiyoshi:2008zw,Nakazato:2008vj,Fischer:2008rh,Nakazato:2013maa}.

{\bf Cosmic core-collapse rates.---}
The cosmic star formation rate history $\dot{\rho}_*(z)$ has become much clearer in recent years.  If every star that forms with a mass $>\,$8$\,M_\odot$ ends with a core collapse, assuming a Salpeter mass function that continues to 100$\,M_\odot$ yields $\dot{n}_{\rm CC}(z)$$\,=\,$$\zeta_{\rm CC}\,\dot{\rho}_*(z)$, with $\zeta_{\rm CC} = 0.0074 / M_{\odot}$, as shown in Fig.~\ref{rates} for both the SFR data compiled in \cite{Hopkins:2006bw} and the parametrized form from \cite{Yuksel:2008cu,Kistler:2011yk}.  (The IMF dependence is small, see \cite{ Hopkins:2006bw}.)

Measurements of the cosmic rate of core-collapse supernovae have also greatly improved.  In \cite{Horiuchi:2011zz}, it was noted that such SN data was lower by a factor of $\sim\,$2 than the inferred $\dot{n}_{\rm CC}(z)$.  In Fig.~\ref{rates}, we see that the latest measurements \cite{Melinder:2012nv,Mattila:2012zr,Dahlen:2012cm} narrow this to a degree, although a gap persists at increasing $z$.  The fraction of core collapses that fail to produce a SN, and thus cannot be counted by SN surveys, remains largely unconstrained.  The existence of stellar-mass black holes, e.g., in binaries \cite{Orosz:2007ng}, at least requires a non-zero black hole birth rate, which may or may not have been accompanied by a visible SN.

A fairly general theoretical expectation is that stars with lower metal content should form more massive cores at the time of collapse (due to brighter burning and lower mass loss over their lifetimes), leading in turn to a higher prevalence of failed explosions and black hole production \cite{Heger:2002by,Zhang:2007nw,O'Connor:2010tk}.  The rate of Type Ib/Ic SNe, which are believed to arise from very massive stars that have lost their envelopes due to metal-line driven winds \cite{Heger:2002by}, may then be suppressed if such stars fail to explode.  The extensive Lick Observatory Supernova Search found that the SN~Ibc to core collapse ratio decreases by a factor of $\sim\,$3 in galaxies at $\lesssim\,$10$^{10}\, M_\odot$ (see Fig.~23 in \cite{Li:2010kd}).  Indeed, galaxies at low redshift show a substantial drop in average metallicity below $\sim\,$10$^{10}\, M_\odot$ \cite{Gallazzi:2005df}.

Since the universe was less metal-enriched at higher redshifts, we pursue an evolving model for the cosmic unnova fraction.  We follow the indications given by bright gamma-ray bursts from \cite{Kistler:2009mv,Kistler:2013jza} of a rate that evolves more strongly than the SFR by a factor of $\sim\,$(1$\,+\,$$z$).  Fig.~\ref{rates} shows an unnova fraction that is 10\% of the total core-collapse rate locally and grows with $z$ ({\it dashed line}).

Fig.~\ref{rates} also displays our expected SN rate ({\it dotted line}).  It is possible that the threshold mass for core collapse itself depends on metallicity, although a decrease below 8$\,M_\odot$ would likely increase the rate of low-luminosity O-Ne-Mg explosions \cite{Poelarends:2007ip,Prieto:2008bw,Thompson:2008sv,Smartt:2008zd,Pumo:2009bg} that may not necessarily increase the observed rate of SNe.  Since corrections of high-redshift data due to incompleteness of such faint events are based on local observations, where the metallicity is highest, we do not attempt an additional correction.

\begin{figure}[t!]
\includegraphics[width=3.35in,clip=true]{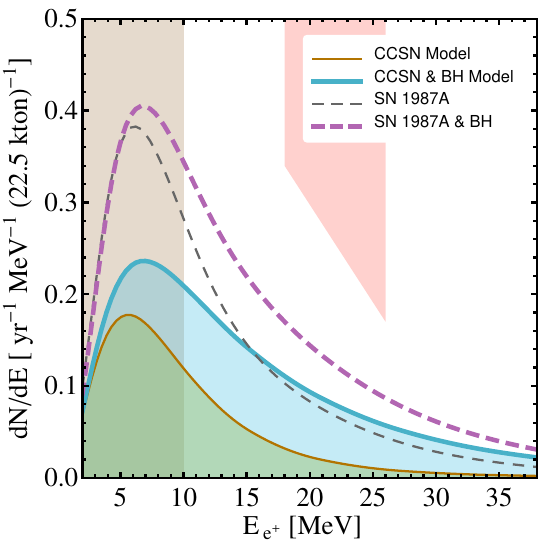}
\caption{The positron production spectrum in 22.5~kton Super-Kamiokande for our cosmic MeV neutrino background models (as labelled).  Denoted are the regions where reactor antineutrinos dominate (below 10~MeV) and the inferred limits (from 18--26~MeV) based on 2003 Super-K data \cite{Malek:2002ns} from \cite{Horiuchi:2008jz} (2012 Super-K limits are model dependent \cite{Bays:2011si}).
}
\label{cmnb}
\end{figure}

\begin{table}[t]
\caption{CMNB event rates in various ranges of visible energy from the spectra displayed in Fig.~\ref{cmnb}. All quoted values are per 22.5 kton~yr (per 0.560 Mton$\,\times\,$10 yr).\label{table:dsnb}}
\begin{ruledtabular}
\begin{tabular}{lccc}
Range (MeV) & {4-10} &{10-18} & {18-26}\\
\hline
CCSN Model	        & 0.95 (238) & 0.54 (135) & 0.14 (36) \\
CCSN \& BH Model    & 1.34 (335) & 1.25 (314) & 0.65 (162) \\
SN$\,$1987A            & 2.09 (523) & 1.40 (350) & 0.56 (139) \\
SN$\,$1987A \& BH	    & 2.26 (566) & 1.97 (492) & 1.00 (249)
\end{tabular}
\end{ruledtabular}
\end{table}

{\bf The Cosmic MeV Neutrino Background.---}
The flux of neutrinos from cosmic core collapses depends on their spectra and rate history, as discussed above, as well as the cosmology assumed.  Including the cross section for inverse-beta decay $\sigma(E_{\bar\nu_e})$ \cite{Vogel:1999zy,Strumia:2003zx}, we obtain the positron spectrum in the detector in terms of $E_{e^+}$$\,=\,$$E_{\bar\nu_e}$$\, - \,$$\Delta$, where $\Delta$$\, = \,$${M}_{n}$$\, - \,$${M}_{p}$, as
\begin{equation} 
  \psi(E_{e^+}) = c\, \sigma(E_{\nu}) N_t \int_{0}^{z_{\rm max}} \frac{dN_{\nu}}{dE_{\nu}^\prime}  \frac{dE_{\nu}^\prime}{dE_{\nu}}\, \frac{\dot{n}(z)}{dz/dt} \,dz \,,\nonumber
\label{eq:dsnb}
\end{equation}
where $dz/dt$$\, = \,$$H_0\, (1+z) [\Omega_m (1+z)^3 +\Omega_\Lambda ]^{1/2}$ (with $\Omega_m$$\, = \,$$0.3$, $\Omega_{\Lambda}$$\, = \,$$0.7$, and ${H}_{0}$$\, = \,$$70\,$km/s/Mpc) and $dE_\nu^\prime/dE_\nu$$\,=\,$$(1+z)$ accounts for redshift.  For a 22.5~kton fiducial volume, such as Super-Kamiokande, $N_t = 1.5 \times 10^{33}$.

In Fig.~\ref{cmnb}, we present the positron spectra obtained from our models discussed above for $dN_{\bar\nu_e}/dE_{\bar\nu_e}$ and $\dot{n}(z)$, in which the contribution is either entirely from SNe ({\it thin solid and dashed lines}; i.e., the typical DSNB) or from a combination of SNe and unnovae as in Fig.~\ref{rates} ({\it thick solid and dashed lines}). In Table~\ref{table:dsnb}, we provide event rates from these models in given energy ranges.

We see that unnovae could contribute more than half of the CMNB in the 10--20 MeV range and easily be the dominant contribution above 20~MeV.  If backgrounds are reduced by the addition of gadolinium~\cite{Beacom:2003nk}, these should be detectable.  The improved capabilities of Super-Kamiokande~IV were recently shown to allow detection of the 2.2~MeV gamma-ray associated with $n+p\rightarrow d + \gamma$ with a $\sim\,$20\% efficiency \cite{Zhang(2012),Chen:2012ji}, already permitting at least partial tagging of inverse-beta events.

\begin{figure}[t!]
\includegraphics[width=3.35in,clip=true]{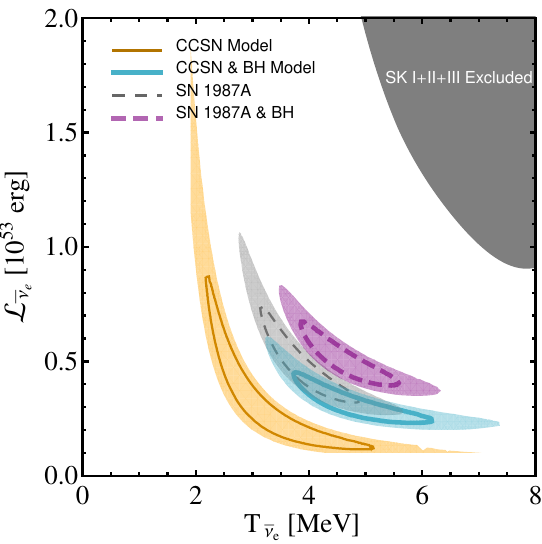}
\caption{Inferred constraints on the combination of SN $\bar\nu_e$ temperature and luminosity expected in a 560~kton detector (i.e., Hyper-Kamiokande) in 10 years if a Fermi-Dirac spectrum was naively assumed.  Shown are the four models from Fig.~\ref{cmnb}; lines (shades) correspond to $2\sigma$ ($5\sigma$) contours.  We see that, if unnovae were not accounted for, the reconstructed properties of ``SNe'' (as exhibited by thick contour sets) would be far from the ``true'' values (thin contour sets).  A Galactic SN, or detailed SN simulations for an ensemble of progenitors, may supersede SN~1987A data to isolate the unnova
component.
}
\label{excl}
\end{figure}

In Fig.~\ref{excl}, we follow the procedure of \cite{Yuksel:2005ae} to determine what would be inferred from the detailed observations of the CMNB afforded by a 560~kton detector such as Hyper-Kamiokande (with Gd).  This imposes a Fermi-Dirac spectrum with cosmic evolution following the star formation rate from Fig.~\ref{rates}.  We reconstruct both $2\sigma$ ({\it lines}) and $5\sigma$ ({\it shaded}) allowed regions in the temperature versus luminosity plane if the observed signal follows one of the four scenarios in Fig.~\ref{cmnb}, using only the 10--20 MeV range in which background should be lowest \cite{Beacom:2003nk}.

We see that, if unnovae are as important as suggested and are not accounted for properly, the inferred SN $\bar\nu_e$ temperature and luminosity would be $\sim\,$5~MeV and $\mathcal{L}_{\bar\nu_e}$$\,\sim \,$3--5$\,\times \,$$10^{52}\,$erg, which are far from the ``true'' values for normal SNe used.  For example, if the SN~1987A model ({\it thin dashed line}) represents the true SN spectrum and the measured CMNB suggests a harder and more energetic spectrum ({\it thick dashed line}), a significant unnovae contribution could be established at $>5\sigma$.
 
Even lower unnova contributions ($\lesssim\,$3\% of the local CC rate with $1+z$ evolution) can be probed provided we have a reliable a priori spectrum for SNe.  If we cannot make such an assumption, it is more challenging to establish the precise contributions solely based on CMNB observations.  This is illustrated in Fig.~\ref{excl} by the overlap in the allowed regions for the SN~1987A ({\it thin dashed}) and CCSN \& BH ({\it thick solid}) models, which shows that a relatively-cold supernova and relatively-hot unnova combination could mimic a signal based on SN~1987A alone.  However, observing a Galactic SN will greatly improve upon the SN~1987A data.

{\bf Discussion and Conclusions.---}
A common question is: ``What can actually be learned from detecting the cosmic MeV neutrino background?''  We have attempted to show an important application.  That core-collapse events occur that produce black holes is inevitable, although the rate remains highly uncertain.  Our first and foremost conclusion is that a significant portion of the CMNB could be due to black holes even if the local rate is measured to be low, due to the metallicity evolution of the universe yielding relatively more unnovae at higher redshifts.  This also affects many other expectations, such as for nucleosynthesis and feedback in young galaxies.  

We have seen that the CMNB is likely detectable in Super-K, even if the neutrino spectrum from SNe is colder than has often been assumed, as with the simulations of \cite{Nakazato:2012qf} that form the DSNB in one of our models, for plausible levels of black hole production.  Other than the possibility of detecting minibursts of neutrino events from core collapses in nearby galaxies with Mton-scale detectors~\cite{Ando:2005ka,Kistler:2008us,Yang:2011xd,Kistler:2012as}, the CMNB provides the only imminent means of testing simulations of the processes occurring deep within dying stars.

It is evident that it may be difficult to determine the average neutrino spectrum from the SNe that form the DSNB, which is subdominant.  However, a new window opens on the study of the formation of black holes.  This is particularly important since the Milky Way is an evolved, metal-enriched galaxy, with an unnova rate that is likely lower than in the distant universe, lowering the odds of directly measuring the spectrum from such an event.

The parametrization of evolution that we use likely saturates
at some redshift.  It would thus be useful to estimate cosmic
unnova rates.  Upcoming surveys, such as LSST \cite{Abell:2009aa},
will detect a large number of SNe that can be compared to the cosmic
star formation rate \cite{Lien:2010yb}.  Determining their
sensitivity to relative evolution, as displayed in Fig.~\ref{rates}, requires
consideration of survey details, which we encourage to be performed,
though is beyond our present scope.

To do so via the difference between SFR and SN data requires consideration of binary interactions amongst massive stars.  In \cite{Sana:2012px}, it was suggested that $\sim\,$25\% of massive O-type stars will be involved in a merger.  As interactions depend on the binary mass ratio and orbital period, any massive stars lost to mergers with a more massive star prior to core collapse would likely be from lower masses.  The SN rate may then be reduced by a factor $f_{\rm m}$, perhaps $\sim\,$5\%.  Mergers could lead to more unnovae, since more high mass stars might be made than merged away, although the net effect of binary interaction is unclear.  This has not been included previously and we defer a more detailed account.

More directly, a 10\% unnova fraction is near the limits of a 10 year ``survey about nothing'' for disappearing massive stars \cite{Kochanek:2008mp}, with some such candidates seen that could indicate a rate near this level \cite{Gerke:2014ooa,Reynolds}.  Further, calculations by \cite{Lovegrove:2013ssa,Piro:2013voa} suggest that a core collapse proceeding to a black hole may yield a distinct cool, but faint optical transient that can be observed \cite{Adams:2013ana}.  Although these techniques are limited to relatively-nearby galaxies, where the unnova fraction should be lower than the more distant universe, we have shown that CMNB measurements can still prove powerful even if such surveys do not find nothing locally.

We note that even if an uncertainty in the overall core-collapse rate causes an overall shift along the $\mathcal{L}$ axis in Fig.~\ref{excl}, temperature information still allows for separation of the relative contributions.  This is important since it is unlikely that we will ever measure the neutrino output of a Galactic unnova, leaving the conditions of black hole formation, and their fraction in the high-redshift universe, purely in the realm of CMNB studies.

When a Galactic SN does occur, this could be used as a detailed template for subtraction of the DSNB, utilizing the measured rate of visible SNe, to arrive at the naked black hole contribution.  Improved knowledge of the nuclear equation of state, which affects the black hole transition and associated neutrino output \cite{Sumiyoshi:2007pp,Fischer:2008rh}, may well be independently obtained \cite{Lattimer(2012)}, further enhancing the extraction of physics from the CMNB.

%
We thank John Beacom and Wick Haxton for comments and discussions.

HY was supported by the LANL LDRD program, during a visit to Berkeley by DOE contract DE-SC00046548 (to WH) and by the Scientific and Technological Research Council of Turkey (TUBITAK), cofunded by Marie Curie Actions under FP7; MDK by NASA through the Einstein Fellowship Program, grant PF0-110074, by Department of Energy contract DE-AC02-76SF00515, and the KIPAC Kavli Fellowship made possible by The Kavli Foundation.


\linespread{0.9}


\vspace{-0.6cm}
\begin{thebibliography}{99}
\vspace{-0.6cm}

\bibitem{Baade} 
  W.~Baade and F.~Zwicky,
  Proc.\ Nat.\ Acad.\ Sci.\ {\bf 20}, 254 (1934).

\bibitem{Colgate:1966ax} 
  S.~A.~Colgate and R.~H.~White,
  Astrophys.\ J.\  {\bf 143}, 626 (1966).

\bibitem{Raffelt:1996wa} 
  G.~G.~Raffelt,
  ``Stars as laboratories for fundamental physics,''
  Univ.\ of Chicago Press, (1996).

\bibitem{Arnett:1996ev} 
  D.~Arnett,
  ``Supernovae and nucleosynthesis,''
  Princeton Univ.\ Press, (1996).

\bibitem{Heger:2002by}
  A.~Heger, C.~L.~Fryer, S.~E.~Woosley, N.~Langer and D.~H.~Hartmann,
  Astrophys.\ J.\  {\bf 591}, 288 (2003).

\bibitem{Marek:2008qi} 
  A.~Marek, H.~T.~.Janka and E.~Mueller,
  Astron.\ Astrophys.\ {\bf 496}, 475 (2009).

\bibitem{Yakunin:2010fn} 
  K.~N.~Yakunin, P.~Marronetti, A.~Mezzacappa {\it et al.},
  Class.\ Quant.\ Grav.\  {\bf 27}, 194005 (2010).

\bibitem{Ott:2012kr} 
  C.~D.~Ott, E.~Abdikamalov, E.~O'Connor, C.~Reisswig, R.~Haas, P.~Kalmus, S.~Drasco, A.~Burrows and E.~Schnetter,
  Phys.\ Rev.\ D {\bf 86}, 024026 (2012).



\bibitem{Melinder:2012nv}
  J.~Melinder {\it et al.},
  Astron.\ Astrophys.\ {\bf 545}, A96 (2012).

\bibitem{Botticella:2011nd} 
  M.~T.~Botticella, S.~J.~Smartt, R.~C.~Kennicutt, {\it et al.},
  Astron.\ Astrophys.\ {\bf 537}, A132 (2012).

\bibitem{Mattila:2012zr} 
  S.~Mattila {\it et al.},
  Astrophys.\ J.\  {\bf 756}, 111 (2012).

\bibitem{Dahlen:2012cm} 
  T.~Dahlen, L.~G.~Strolger, A.~G.~Riess {\it et al.},
  Astrophys.\ J.\  {\bf 757}, 70 (2012).

\bibitem{Li:2010kd} 
  W.~Li, R.~Chornock, J.~Leaman, A.~V.~Filippenko {\it et al.},
  Mon.\ Not.\ Roy.\ Astron.\ Soc.\ {\bf 412}, 1473 (2011).

\bibitem{Graur:2011cv} 
  O.~Graur, D.~Poznanski, D.~Maoz {\it et al.},
  Mon.\ Not.\ Roy.\ Astron.\ Soc.\  {\bf 417}, 916 (2011).

\bibitem{Bazin:2009mp} 
  G.~Bazin, N.~Palanque-Delabrouille, J.~Rich, {\it et al.},
  Astron.\ Astrophys.\ {\bf 499}, 653 (2009).

\bibitem{Cappellaro:1999qy} 
  E.~Cappellaro, R.~Evans and M.~Turatto,
  Astron.\ Astrophys.\  {\bf 351}, 459 (1999).

\bibitem{Horiuchi:2011zz} 
  S.~Horiuchi {\it et al.},
  Astrophys.\ J.\  {\bf 738}, 154 (2011).

\bibitem{Hopkins:2006bw}
  A.~M.~Hopkins and J.~F.~Beacom,
  Astrophys.\ J.\  {\bf 651}, 142 (2006).


\bibitem{Fender:2012tx} 
  R.~Fender and T.~Belloni,
  Science {\bf 337}, 540 (2012).

\bibitem{Mirabel:2003st} 
  F.~Mirabel and I.~Rodrigues,
  Science {\bf 300}, 1119 (2003).

\bibitem{Oppenheimer:1939ue} 
  J.~R.~Oppenheimer and H.~Snyder,
  Phys.\ Rev.\  {\bf 56}, 455 (1939).

\bibitem{Wilson(1971)}
  J.~R.~Wilson,
  Astrophys.\ J.\   {\bf 163}, 209 (1971).

\bibitem{Fryer:1999mi} 
  C.~L.~Fryer,
  Astrophys.\ J.\  {\bf 522}, 413 (1999).

\bibitem{Zhang:2007nw} 
  W.~Q.~Zhang, S.~E.~Woosley and A.~Heger,
  Astrophys.\ J.\  {\bf 679}, 639 (2008).

\bibitem{O'Connor:2010tk} 
  E.~O'Connor and C.~D.~Ott,
  Astrophys.\ J.\  {\bf 730}, 70 (2011).

\bibitem{Ugliano:2012kq} 
  M.~Ugliano, H.~T.~Janka, A.~Marek and A.~Arcones,
  Astrophys.\ J.\  {\bf 757}, 69 (2012).

\bibitem{Horiuchi:2014ska} 
  S.~Horiuchi, K.~Nakamura, T.~Takiwaki, K.~Kotake and M.~Tanaka,
  MNRASL 445, L99-L103 (2014)
  [arXiv:1409.0006 [astro-ph.HE]].

\bibitem{Kochanek:2008mp} 
  C.~S.~Kochanek {\it et al.},
  Astrophys.\ J.\  {\bf 684}, 1336 (2008).

\bibitem{Zwicky:1940zz} 
  F.~Zwicky,
  Rev.\ Mod.\ Phys.\  {\bf 12}, 66 (1940).


\bibitem{Burrows:1988ba} 
  A.~Burrows,
  Astrophys.\ J.\  {\bf 334}, 891 (1988).

\bibitem{Baumgarte:1996iu} 
  T.~W.~Baumgarte, S.~A.~Teukolsky, S.~L.~Shapiro, H.~T.~Janka and W.~Keil,
  Astrophys.\ J.\  {\bf 468}, 823 (1996).

\bibitem{Sumiyoshi:2007pp} 
  K.~Sumiyoshi, S.~Yamada and H.~Suzuki,
  Astrophys.\ J.\  {\bf 667}, 382 (2007).

\bibitem{Nakazato:2008vj} 
  K.~Nakazato, K.~Sumiyoshi, H.~Suzuki and S.~Yamada,
  Phys.\ Rev.\ D {\bf 78}, 083014 (2008).

\bibitem{Sumiyoshi:2008zw} 
  K.~Sumiyoshi, S.~Yamada and H.~Suzuki,
  Astrophys.\ J.\   {\bf 688}, 1176, (2008).

\bibitem{Fischer:2008rh} 
  T.~Fischer, S.~C.~Whitehouse, A.~Mezzacappa, F.~K.~Thielemann and M.~Liebendorfer,
  Astron.\ Astrophys.\ {\bf 499}, 1 (2009).

\bibitem{Nakazato:2013maa} 
  K.~i.~Nakazato,
  Phys.\ Rev.\ D {\bf 88}, no. 8, 083012 (2013)
  [arXiv:1306.4526 [astro-ph.HE]].

\bibitem{Nakazato:2012qf}
  K.~Nakazato, K.~Sumiyoshi, H.~Suzuki, T.~Totani, H.~Umeda and S.~Yamada,
  Astrophys.\ J.\ Supp.\  {\bf 205}, 2 (2013).


\bibitem{Nakazato:2015rya} 
  K.~Nakazato, E.~Mochida, Y.~Niino and H.~Suzuki,
  arXiv:1503.01236 [astro-ph.HE].




\bibitem{ZG}
  Ya.~B.~Zel'dovich and O.~Kh.~Guseinov,
  Sov.\ Phys.\ Dokl.\, {\bf 10}, 524 (1965).
  
\bibitem{Hartmann:1997qe}
  D.~H.~Hartmann and S.~E.~Woosley,
  Astropart.\ Phys.\  {\bf 7}, 137 (1997).

\bibitem{Kaplinghat:1999xi}
  M.~Kaplinghat, G.~Steigman and T.~P.~Walker,
  Phys.\ Rev.\ D {\bf 62}, 043001 (2000).

\bibitem{Ando:2002ky}
  S.~Ando, K.~Sato and T.~Totani,
  Astropart.\ Phys.\  {\bf 18}, 307 (2003).

\bibitem{Strigari:2005hu} 
  L.~E.~Strigari, J.~F.~Beacom, T.~P.~Walker and P.~Zhang,
  JCAP {\bf 0504}, 017 (2005).

\bibitem{Yuksel:2005ae}
  H.~Yuksel, S.~Ando and J.~F.~Beacom,
  Phys.\ Rev.\ C {\bf 74}, 015803 (2006).

\bibitem{Yuksel:2007mn}
  H.~Yuksel and J.~F.~Beacom,
  Phys.\ Rev.\  D {\bf 76}, 083007 (2007).

\bibitem{Horiuchi:2008jz} 
  S.~Horiuchi, J.~F.~Beacom and E.~Dwek,
  Phys.\ Rev.\ D {\bf 79}, 083013 (2009).


\bibitem{Beacom:2010kk} 
  J.~F.~Beacom,
  Ann.\ Rev.\ Nucl.\ Part.\ Sci.\  {\bf 60}, 439 (2010).

\bibitem{Lunardini:2010ab} 
  C.~Lunardini,
  arXiv:1007.3252.


\bibitem{Lunardini:2009ya} 
  C.~Lunardini,
  Phys.\ Rev.\ Lett.\  {\bf 102}, 231101 (2009).

\bibitem{Lien:2010yb} 
  A.~Lien, B.~D.~Fields and J.~F.~Beacom,
  Phys.\ Rev.\ D {\bf 81}, 083001 (2010).

\bibitem{Keehn:2010pn} 
  J.~G.~Keehn and C.~Lunardini,
  Phys.\ Rev.\ D {\bf 85}, 043011 (2012).


\bibitem{Tremonti:2004et} 
  C.~A.~Tremonti, {\it et al.},
  Astrophys.\ J.\  {\bf 613}, 898 (2004).

\bibitem{Kewley}
  L.~Kewley and H.~A.~Kobulnicky,
  in Starbursts: From 30 Doradus to Lyman Break Galaxies {\bf 329}, 307 (2005).

\bibitem{Erb:2006qy} 
  D.~K.~Erb, A.~E.~Shapley, M.~Pettini, C.~C.~Steidel, N.~A.~Reddy and K.~L.~Adelberger,
  Astrophys.\ J.\  {\bf 644}, 813 (2006).

\bibitem{Savaglio:2005hi} 
  S.~Savaglio, {\it et al.},
  Astrophys.\ J.\  {\bf 635}, 260 (2005).

\bibitem{Maiolino:2008gh} 
  R.~Maiolino, {\it et al.},
  Astron.\ Astrophys.\  {\bf 488}, 463 (2008).


\bibitem{Woosley:1993wj} 
  S.~E.~Woosley,
  Astrophys.\ J.\  {\bf 405}, 273 (1993).

\bibitem{Metzger:2010pp} 
  B.~D.~Metzger, D.~Giannios, T.~A.~Thompson, N.~Bucciantini and E.~Quataert,
  Mon.\ Not.\ Roy.\ Astron.\ Soc.\  {\bf 413}, 2031 (2011).

\bibitem{Stanek:2006gc} 
  K.~Z.~Stanek {\it et al.},
  Acta Astron.\  {\bf 56}, 333 (2006).

\bibitem{Graham:2012ga} 
  J.~F.~Graham and A.~S.~Fruchter,
  Astrophys.\ J.\  {\bf 774}, 119 (2013).

\bibitem{Yoon:2005tv} 
  S.~C.~Yoon and N.~Langer,
  Astron.\ Astrophys.\  {\bf 443}, 643 (2005).

\bibitem{Woosley:2005gy} 
  S.~Woosley and A.~Heger,
  Astrophys.\ J.\  {\bf 637}, 914 (2006).



\bibitem{Daigne:2006kf} 
  F.~Daigne, E.~M.~Rossi and R.~Mochkovitch,
  Mon.\ Not.\ Roy.\ Astron.\ Soc.\  {\bf 372}, 1034 (2006).

\bibitem{Le:2006pt} 
  T.~Le and C.~D.~Dermer,
  Astrophys.\ J.\  {\bf 661}, 394 (2007).

\bibitem{Yuksel:2006qb}
  H.~Yuksel and M.~D.~Kistler,
  Phys.\ Rev.\ D {\bf 75}, 083004 (2007).

\bibitem{Kistler:2007ud} 
  M.~D.~Kistler, H.~Yuksel, J.~F.~Beacom and K.~Z.~Stanek,
  Astrophys.\ J.\  {\bf 673}, L119 (2008).

\bibitem{Kistler:2009mv} 
  M.~D.~Kistler, H.~Yuksel, J.~F.~Beacom, A.~M.~Hopkins and J.~S.~B.~Wyithe,
  Astrophys.\ J.\  {\bf 705}, L104 (2009).

\bibitem{Butler:2009nx} 
  N.~R.~Butler, J.~S.~Bloom and D.~Poznanski,
  Astrophys.\ J.\  {\bf 711}, 495 (2010).

\bibitem{Kistler:2013jza} 
  M.~D.~Kistler, H.~Yuksel and A.~M.~Hopkins,
  arXiv:1305.1630.



\bibitem{Lovegrove:2013ssa} 
  E.~Lovegrove and S.~E.~Woosley,
  Astrophys.\ J.\  {\bf 769}, 109 (2013).

\bibitem{Piro:2013voa} 
  A.~L.~Piro,
  Astrophys.\ J.\  {\bf 768}, L14 (2013).



\bibitem{Malek:2002ns}
  M.~Malek {\it et al.},
  Phys.\ Rev.\ Lett.\  {\bf 90}, 061101 (2003).

\bibitem{Eguchi:2003gg}
  K.~Eguchi {\it et al.},
  Phys.\ Rev.\ Lett.\  {\bf 92}, 071301 (2004).

\bibitem{Aharmim:2006wq}
  B.~Aharmim {\it et al.},
  Astrophys.\ J.\  {\bf 653}, 1545 (2006).

\bibitem{Bays:2011si}
  K.~Bays {\it et al.},
  Phys.\ Rev.\ D {\bf 85}, 052007 (2012).



\bibitem{Abe:2011ts} 
  K.~Abe {\it et al.},
  arXiv:1109.3262.

\bibitem{Goon:2012if} 
  J.~Goon {\it et al.},
  arXiv:1204.2295.

\bibitem{Autiero:2007zj} 
  D.~Autiero {\it et al.},
  JCAP {\bf 0711}, 011 (2007).


%
\bibitem{Abell:2009aa} 
  P.~A.~Abell {\it et al.}  [LSST Science and LSST Project Collaborations],
  arXiv:0912.0201.


 \bibitem{Hirata:1987hu}
  K.~Hirata {\it et al.},
  Phys.\ Rev.\ Lett.\  {\bf 58}, 1490 (1987).

\bibitem{Bionta:1987qt}
  R.~M.~Bionta {\it et al.},
  Phys.\ Rev.\ Lett.\  {\bf 58}, 1494 (1987).



\bibitem{Chakraborty:2008zp} 
  S.~Chakraborty, S.~Choubey, B.~Dasgupta and K.~Kar,
  JCAP {\bf 0809}, 013 (2008).

\bibitem{Galais:2009wi} 
  S.~Galais, J.~Kneller, C.~Volpe and J.~Gava,
  Phys.\ Rev.\ D {\bf 81}, 053002 (2010).

\bibitem{Duan:2010bg}
  H.~Duan, G.~M.~Fuller and Y.~Z.~Qian,
  Ann.\ Rev.\ Nucl.\ Part.\ Sci.\  {\bf 60}, 569 (2010).

\bibitem{Lunardini:2012ne} 
  C.~Lunardini and I.~Tamborra,
  JCAP {\bf 1207}, 012 (2012).



\bibitem{Yuksel:2008cu} 
  H.~Yuksel, M.~D.~Kistler, J.~F.~Beacom and A.~M.~Hopkins,
  Astrophys.\ J.\  {\bf 683}, L5 (2008).

\bibitem{Kistler:2011yk} 
  M.~D.~Kistler {\it et al.},
  Astrophys.\ J.\  {\bf 770}, 88 (2013).


\bibitem{Orosz:2007ng} 
  J.~A.~Orosz {\it et al.},
  Nature {\bf 449}, 872 (2007).


\bibitem{Gallazzi:2005df} 
  A.~Gallazzi, S.~Charlot, J.~Brinchmann, S.~D.~M.~White and C.~A.~Tremonti,
  Mon.\ Not.\ Roy.\ Astron.\ Soc.\  {\bf 362}, 41 (2005).


\bibitem{Poelarends:2007ip}
  A.~J.~T.~Poelarends, F.~Herwig, N.~Langer and A.~Heger,
  Astrophys.\ J.\  {\bf 675}, 614 (2008).

\bibitem{Prieto:2008bw} 
  J.~L.~Prieto {\it et al.},
  Astrophys.\ J.\  {\bf 681}, L9 (2008).

\bibitem{Thompson:2008sv} 
  T.~A.~Thompson {\it et al.},
  Astrophys.\ J.\  {\bf 705}, 1364 (2009).

\bibitem{Smartt:2008zd}
  S.~J.~Smartt {\it et al.},
  Mon.\ Not.\ Roy.\ Astron.\ Soc.\  {\bf 395}, 1409 (2009).

\bibitem{Pumo:2009bg} 
  M.~L.~Pumo {\it et al.},
  Astrophys.\ J.\  {\bf 705}, L138 (2009).






\bibitem{Vogel:1999zy}
  P.~Vogel and J.~F.~Beacom,
  Phys.\ Rev.\  D {\bf 60}, 053003 (1999).

\bibitem{Strumia:2003zx} 
  A.~Strumia and F.~Vissani,
  Phys.\ Lett.\ B {\bf 564}, 42 (2003).



\bibitem{Beacom:2003nk}
  J.~F.~Beacom and M.~R.~Vagins,
  Phys.\ Rev.\ Lett.\  {\bf 93}, 171101 (2004).
    
\bibitem{Zhang(2012)}
  H.~Zhang, Ph.D.\ thesis, (University of Tsinghua, 2012).
  %

\bibitem{Chen:2012ji} 
  S.~Chen,
  J.\ Phys.\ Conf.\ Ser.\  {\bf 375}, 042051 (2012).


\bibitem{Ando:2005ka}
  S.~Ando, J.~F.~Beacom and H.~Yuksel,
  Phys.\ Rev.\ Lett.\  {\bf 95}, 171101 (2005).

\bibitem{Kistler:2008us} 
  M.~D.~Kistler, H.~Yuksel, S.~Ando, J.~F.~Beacom and Y.~Suzuki,
  Phys.\ Rev.\ D {\bf 83}, 123008 (2011).

\bibitem{Yang:2011xd} 
  L.~Yang and C.~Lunardini,
  Phys.\ Rev.\ D {\bf 84}, 063002 (2011).

\bibitem{Kistler:2012as} 
  M.~D.~Kistler, W.~C.~Haxton and H.~Yuksel,
  Astrophys.\ J.\  {\bf 778}, 81 (2013).


\bibitem{Sana:2012px} 
  H.~Sana, S.~E.~de Mink, A.~de Koter {\it et al.},
  Science {\bf 337}, 444 (2012).

\bibitem{Gerke:2014ooa} 
  J.~R.~Gerke, C.~S.~Kochanek and K.~Z.~Stanek,
  Mon.\ Not.\ Roy.\ Astron.\ Soc.\  {\bf 450}, 3289 (2015).

\bibitem{Reynolds}
  T.~M.\ Reynolds,  M.\ Fraser, and G.\ Gilmore, 
  Mon.\ Not.\ Roy.\ Astron.\ Soc.\ {\bf 453}, 2885 (2015).

\bibitem{Adams:2013ana} 
  S.~M.~Adams, C.~S.~Kochanek, J.~F.~Beacom, M.~R.~Vagins and K.~Z.~Stanek,
  arXiv:1306.0559.

\bibitem{Lattimer(2012)}
  J.~M.~Lattimer,
  Ann.\ Rev.\ Nucl.\ Part.\ Sci.\ {\bf 62}, 485 (2012).

\end{thebibliography}
\end{document}